\pdfoutput=1

\documentclass[11pt,twoside,a4paper,cmspaper,final,collab]{cms-tdr}
\def\svnVersion{41869:42119M}\def\cmsCernNo{2011-015}\def\cmsCernDate{\today}\def\cmsMessage{Submitted to Physics Letters B\\ \vspace{20pt}\textnormal{Dedicated to the memory of Ulrich Baur}}

\begin{document}\cmsNoteHeader{EWK-10-009}

\hyphenation{had-ron-i-za-tion}
\hyphenation{cal-or-i-me-ter}
\hyphenation{de-vices}
\RCS$Revision: 42119 $
\RCS$HeadURL: svn+ssh://alverson@svn.cern.ch/reps/tdr2/papers/EWK-10-009/trunk/EWK-10-009.tex $
\RCS$Id: EWK-10-009.tex 42119 2011-02-26 16:18:01Z alverson $
\newcommand{\CLs}{\ensuremath{CL_\mathrm{s}}}
\newcommand{\CLb}{\ensuremath{CL_\mathrm{b}}}
\newcommand{\CLsb}{\ensuremath{CL_\mathrm{s+b}}}

\newcommand{\nanob}{\mbox{{\rm ~nb}~}}
\newcommand{\fb}{\ensuremath{\mathrm{fb}}}
\newcommand{\pb}{\ensuremath{\mathrm{pb}}}
\newcommand{\ifb}{\ensuremath{\mathrm{fb^{-1}}}}
\newcommand{\ipb}{\ensuremath{\mathrm{pb^{-1}}}}
\newcommand{\grad}{\ensuremath{^{\circ}}}
\newcommand{\lsim}{\raisebox{-1.5mm}{$\:\stackrel{\textstyle{<}}{\textstyle{\sim}}\:$}}
\newcommand{\gsim}{\raisebox{-1.5mm}{$\:\stackrel{\textstyle{>}}{\textstyle{\sim}}\:$}}

\newcommand{\pipm}{\ensuremath{\pi^{\pm}}}
\newcommand{\pizero}{\ensuremath{\pi^{0}}}
\newcommand{\Hi}{\ensuremath{\mathrm{H}}}
\newcommand{\W}{\ensuremath{\mathrm{W}}}
\newcommand{\Wjets}{\ensuremath{\mathrm{W+jets}}}
\newcommand{\Zjets}{\ensuremath{\mathrm{Z+jets}}}
\newcommand{\Wt}{\ensuremath{\mathrm{Wt}}}
\newcommand{\Wstar}{\ensuremath{\mathrm{W}^{*}}}
\newcommand{\Wparenthesisstar}{\ensuremath{\mathrm{W}^{(*)}}}
\newcommand{\WW}{\ensuremath{\W^+\W^-}}
\newcommand{\Zstar}{\ensuremath{\mathrm{Z}^{*}}}
\newcommand{\ZZ}{\ensuremath{\Z\Z}}
\newcommand{\WZ}{\ensuremath{\W\Z}}
\newcommand{\E}{\ensuremath{\mathrm{e}}}
\newcommand{\Ep}{\ensuremath{\mathrm{e}^{+}}}
\renewcommand{\Em}{\ensuremath{\mathrm{e}^{-}}}
\newcommand{\Epm}{\ensuremath{\mathrm{e}^{\pm}}}
\newcommand{\Emp}{\ensuremath{\mathrm{e}^{\mp}}}
\newcommand{\M}{\ensuremath{\mu}}
\newcommand{\Mp}{\ensuremath{\mu^{+}}}
\newcommand{\Mm}{\ensuremath{\mu^{-}}}
\newcommand{\Mpm}{\ensuremath{\mu^{\pm}}}
\newcommand{\Mmp}{\ensuremath{\mu^{\mp}}}
\newcommand{\Tau}{\ensuremath{\tau}}
\newcommand{\Nu}{\ensuremath{\nu}}
\newcommand{\Nubar}{\ensuremath{\bar{\nu}}}
\newcommand{\Lep}{\ensuremath{\ell}}
\newcommand{\Lepp}{\ensuremath{\ell^{+}}}
\newcommand{\Lepm}{\ensuremath{\ell^{-}}}
\newcommand{\Lprime}{\ensuremath{\Lep^{\prime}}}
\newcommand{\Prot}{\ensuremath{\mathrm{p}}}
\newcommand{\Pbar}{\ensuremath{\bar{\mathrm{p}}}}
\newcommand{\PP}{\Prot\Prot}
\newcommand{\PPbar}{\Prot\Pbar}
\newcommand{\qq}{\ensuremath{\mathrm{q}\mathrm{q}}}
\newcommand{\Wtb}{\ensuremath{\W\mathrm{t}\mathrm{b}}}
\newcommand{\Atop}{\ensuremath{\bar{\mathrm{t}}}}
\newcommand{\Abot}{\ensuremath{\bar{\mathrm{b}}}}
\newcommand{\To}{\ensuremath{\rightarrow}}

\newcommand{\mHi}{\ensuremath{m_{\mathrm{H}}}}
\newcommand{\mW}{\ensuremath{m_{\mathrm{W}}}}
\newcommand{\mZ}{\ensuremath{m_{\mathrm{Z}}}}
\newcommand{\mll}{\ensuremath{m_{\Lep\Lep}}}

\newcommand{\ptveto}{\ensuremath{\pt^\mathrm{veto}}}
\newcommand{\ptl}{\ensuremath{p_\perp^{\Lep}}}
\newcommand{\ptlmax}{\ensuremath{p_{\mathrm{T}}^{\Lep,\mathrm{max}}}}
\newcommand{\ptlmin}{\ensuremath{p_{\mathrm{T}}^{\Lep,\mathrm{min}}}}
\newcommand{\met}{\ensuremath{\Et^{\mathrm{miss}}}}
\newcommand{\delphill}{\ensuremath{\Delta\phi_{\Lep\Lep}}}
\newcommand{\deletall}{\ensuremath{\Delta\eta_{\Lep\Lep}}}
\newcommand{\delphimetl}{\ensuremath{\Delta\phi_{\met\Lep}}}
\newcommand{\Et}{\ensuremath{E_\mathrm{T}}}
\newcommand{\delR}{\ensuremath{\Delta R}}
\newcommand{\Eta}{\ensuremath{\eta}}

\newcommand{\effsig}{\ensuremath{\varepsilon_{\mathrm{bkg}}^{\mathrm{S}}}}
\newcommand{\effnorm}{\ensuremath{\varepsilon_{\mathrm{bkg}}^{\mathrm{N}}}}
\newcommand{\Nsig}{\ensuremath{N_{\mathrm{bkg}}^{\mathrm{S}}}}
\newcommand{\Nnorm}{\ensuremath{N_{\mathrm{bkg}}^{\mathrm{N}}}}

\newcommand{\dyee}{\ensuremath{Z/\gamma^*\to e^+e^-}}
\newcommand{\dymm}{\ensuremath{Z/\gamma^*\to\mu^+\mu^-}}
\newcommand{\dytt}{\ensuremath{Z/\gamma^*\to\tau^+\tau^-}}
\newcommand{\dyll}{\ensuremath{Z/\gamma^*\to\ell^+\ell^-}}
\newcommand{\zee}{\ensuremath{Z\to e^+e^-}}
\newcommand{\zmm}{\ensuremath{Z\to\mu^+\mu^-}}
\newcommand{\ztt}{\ensuremath{Z\to\tau^+\tau^-}}
\newcommand{\zll}{\ensuremath{Z\to\ell^+\ell^-}}
\newcommand{\ppww}{\ensuremath{pp \to W^+W^-}}
\newcommand{\wwll}{\ensuremath{WW\to \ell^+\ell^-}}
\newcommand{\wwlnln}{\ensuremath{W^+W^-\to \ell^+\nu \ell^-\bar{\nu}}}
\newcommand{\ww}{\ensuremath{\mathrm{WW}}}
\newcommand{\wwpm}{\ensuremath{W^+W^-}}
\newcommand{\hww}{\ensuremath{\mathrm{H}\to \mathrm{W}^+\mathrm{W}^-}}
\newcommand{\wz}{\ensuremath{WZ}}
\newcommand{\zz}{\ensuremath{ZZ}}
\newcommand{\wgamma}{\ensuremath{W\gamma}}
\newcommand{\wjets}{\ensuremath{W+}jets}
\newcommand{\tw}{\ensuremath{\mathrm{tW}}}
\newcommand{\singletopt}{\ensuremath{t} ($t$-chan)}
\newcommand{\singletops}{\ensuremath{t} ($s$-chan)}

\def\fixme{({\bf FixMe})}
\newcommand{\ee}{\ensuremath{ee}}
\newcommand{\emu}{\ensuremath{e\mu}}
\def\mm{\ensuremath{\mu\mu}}

\newcommand{\intlumi}{36~\pbinv}

\cmsNoteHeader{EWK-10-009} 
\title{Measurement of W$^+$W$^-$ Production and Search for the Higgs Boson in pp Collisions at $\sqrt{s}$ = 7 TeV}

\address[cern]{CERN}
\author[cern]{The CMS Collaboration}

\date{\today}

\abstract{
A measurement of $\ensuremath{\mathrm{W^+}}\ensuremath{\mathrm{W^-}}$
production in pp collisions at $\sqrt{s} = $ 7~TeV and a search for
the Higgs boson are reported. The $\ensuremath{\mathrm{W^+}}\ensuremath{\mathrm{W^-}}$
candidates are selected in events with two leptons, either electrons or muons.
The measurement is performed using LHC data
recorded with the CMS detector, corresponding to an integrated
luminosity of 36~pb$^{-1}$.
The pp $\to$ $\ensuremath{\mathrm{W^+}}\ensuremath{\mathrm{W^-}}$ cross section is measured to be
$41.1 \pm 15.3 \,(\rm{stat}) \pm 5.8 \,(\rm{syst}) \pm 4.5 \,(\rm{lumi})\,\rm{pb}$,
consistent with the standard model prediction. Limits on
$\ensuremath{\mathrm{W}}\ensuremath{\mathrm{W}}\gamma$ and
$\ensuremath{\mathrm{W}}\ensuremath{\mathrm{W}}\ensuremath{\mathrm{Z}}$ anomalous triple gauge couplings are set.
The search for the standard model Higgs boson in the $\ensuremath{\mathrm{W^+}}\ensuremath{\mathrm{W^-}}$ decay
mode does not reveal any evidence of excess above backgrounds. Limits
are set on the production of the Higgs boson in the context of the standard model and
in the presence of a sequential fourth family of fermions with
high masses. In the latter context, a Higgs boson with mass between 144 and
207~$\ensuremath{\,\mathrm{Ge\kern -0.1em V\!/c}^2}$ is ruled out at
95\% confidence level.  }

\hypersetup{%
pdfauthor={CMS Collaboration},%
pdftitle={Measurement of W+W- Production and Search for the Higgs Boson in pp Collisions at sqrt(s) = 7 TeV},%
pdfsubject={CMS},%
pdfkeywords={CMS, physics, software, computing}}

\maketitle 

 \section{Introduction}
 \label{sec:intro}

The standard model (SM) of particle physics successfully describes
the majority of high-energy experimental data~\cite{pdg}. One of the key
remaining questions is the origin of the masses of $\W$~and $\Z$~bosons.
In the SM, it is attributed to the spontaneous breaking of electroweak
symmetry caused by a new scalar field~\cite{Higgs1, Higgs2, Higgs3}. The
existence of the associated field quantum,
the Higgs boson, has yet to be experimentally confirmed.
The $\WW$ channel is particularly sensitive for the Higgs boson
searches in the intermediate mass range (120 -- 200 $\GeVcc$).

Direct searches at the CERN LEP collider have set a limit on the SM Higgs boson
mass of $m_{\rm{H}}~>$ 114.4 $\GeVcc$ at 95\% confidence level (C.L.)~\cite{LEPHIGGS}.
Precision electroweak measurements constrain the
mass of the SM Higgs boson to be less than 185 $\GeVcc$ at 95\% C.L.~\cite{EWK}.
Direct searches at the Tevatron exclude the SM Higgs boson
in the mass range 158 -- 175 $\GeVcc$ at 95\% C.L.~\cite{TEVHIGGS_ICHEP2010}.

A possible extension to the SM is the addition of a fourth family of fermions~\cite{4thftheory,4thftheory_2}. For
sufficiently large lepton and quark masses, this extension has not been excluded by existing constraints. The
presence of another fermion family produces an enhancement of the dominant gluon fusion
cross section, together with some changes in the Higgs decay branching
fractions. The choice of infinitely heavy quarks of the fourth family in the
extended SM yields to the smallest enhancement factor for the Higgs boson cross section, hence to
the most conservative scenario for the exclusion of such a model.
This scenario is used to set limits in this paper.

The dominant irreducible background for $\hww$ production is the SM nonresonant production of $\WW$.
A good understanding of this process and its properties is thus needed for the Higgs boson search.
The $\WW$ production has been extensively studied by the LEP and Tevatron
experiments~\cite{DELPHI,ALEPH,OPAL,L3,CDF,D0}, where it has been found to be in
agreement with the SM prediction. In $\rm{pp}$ collisions at the LHC, the SM
next-to-leading order (NLO) QCD prediction of the $\WW$ production cross section at
$\sqrt{s} = 7~\TeV$ is $43.0 \pm 2.0$ pb~\cite{MCFM}. The $\WW$ production rates and
differential cross sections are also sensitive to anomalous WW$\gamma$ and WWZ triple
gauge boson couplings (TGC)~\cite{TGC1, TGC2}.

The first measurement of the $\WW$ cross section in pp collisions
at $\sqrt{s} = $ 7~TeV is reported here together with the results of the
related search for the Higgs boson in the
$\WW$ decay mode. The measurement is performed with data corresponding to an integrated
luminosity of 35.5~$\pm$~3.9~pb$^{-1}$, recorded with the
Compact Muon Solenoid (CMS) detector, .
The $\WW$ candidates, with both W bosons decaying leptonically,
are selected in final states consisting of two isolated, high transverse momentum ($\pt$),
oppositely-charged leptons (electrons or muons), and large missing transverse
energy due to the undetected neutrinos. The search for the Higgs boson is
performed in the 120 to 600 $\GeVcc$ mass range, using both a
cut-based event selection and a multivariate analysis. The search results
are interpreted for both a SM Higgs boson and in the presence of a fourth family of fermions.

The paper is organized as follows. Section~\ref{sec:cms} briefly describes the main
components of the CMS detector used in this analysis. Section~\ref{sec:ww} describes
the $\WW$ production cross section measurement. The extraction of the limits
on anomalous TGC is discussed in Section~\ref{sec:aTGCs}.
The $\hww$ search procedure and results are presented in Section~\ref{sec:hww}.

\section{CMS Detector and Simulations}
\label{sec:cms}

The CMS detector is described in detail elsewhere~\cite{CMSdetector}, while the
key components for this analysis are summarized here. The central part of the
CMS detector is a superconducting solenoid, which provides an axial magnetic
field of 3.8~T parallel to the beam axis. Charged particle trajectories
are measured by the silicon pixel and strip tracker, which covers the
pseudorapidity region
$|\eta| < 2.5$. Here, the pseudorapidity is defined as $\eta=-\ln{\tan{\theta/2}}$,
where $\theta$ is the polar angle of the trajectory of the particle
with respect to the direction of the counterclockwise beam. A crystal electromagnetic
calorimeter (ECAL) and a brass/scintillator hadron calorimeter (HCAL)
surround the tracking volume and cover $|\eta| < 3.0$. A quartz-fiber
Cherenkov calorimeter (HF) extends the coverage to $|\eta| = 5.0$.
Muons are measured in gas detectors embedded in the iron return yoke outside the
solenoid, in the pseudorapidity range  $|\eta|<2.4$.
The detector is nearly hermetic, allowing for energy
balance measurements in the plane transverse to the beam axis.

The first level of the trigger system, composed of custom
hardware processors, is designed to select the most interesting events
in less than $1~\mathrm{ms}$, using information from the calorimeters and muon
detectors. The High Level Trigger processor farm further
decreases the event rate to a few hundred Hz, before data storage.

For this analysis, the $\hww$ and Drell--Yan processes are generated
with the {\sc POWHEG} program~\cite{powheg}. The $\rm{q}\bar{\rm{q}}\to\WW$, $\Wjets$,
t$\bar{\rm{t}}$ and the \tw\ processes are generated with the {\sc
MADGRAPH} event generator~\cite{madgraph}, the gg$\to\WW$ process is
simulated with the {\sc GG2WW} event generator~\cite{ggww}, and the
remaining processes are generated with {\sc
PYTHIA}~\cite{pythia}. A set of parton distribution functions (PDF)
used for the simulated samples is CTEQ6L~\cite{cteq66}.
Calculations at next-to-next-to-leading order (NNLO) are
used for the $\hww$ process, while NLO calculations are used for
background cross sections. All processes are simulated using a detailed
description of the CMS detector, based on the {\sc GEANT4}
package~\cite{Agostinelli:2002hh}.

\section{Standard Model $\WW$ Cross Section Measurement}
\label{sec:ww}

\subsection{Event selection}
\label{subsec:ww_evtsel}

Several SM processes can lead to a reconstructed final state similar to that of the $\WW$ signal.
These backgrounds include instrumental contributions from $\Wjets$ and ${\rm QCD}$
multijet events where at least one of the jets is mis-identified as a lepton,
top quark production (t$\bar{\rm{t}}$ and tW), the Drell--Yan $\dyll$ process, and diboson
production (W$\gamma$, $\WZ$ and $\ZZ$).

Events are selected with two high-$\pt$, oppositely-charged isolated leptons,
in three final states: $\Ep\Em$, $\Mp\Mm$ and
$\Epm\Mmp$. These final states thus include $\W\to\tau\nu_{\tau}$
events with leptonic $\tau$ decays. The online event trigger requires the
presence of a high-$\pt$ electron or muon~\cite{wzxs}. The trigger efficiency for signal
events, which would be selected by the full offline event selection,
is found to be above 98\% in the $\Mp\Mm$ final state and above
99\% in the $\Ep\Em$ and $\Epm\Mmp$ final states.

Muon candidates are reconstructed combining two algorithms~\cite{muonpas},
one in which tracks in the silicon detector are matched to hits in the
muon system, and another in which a global fit is performed on hits
in both the silicon tracker and the muon system. All muon
candidates are required to be successfully reconstructed by both algorithms and to
have $\pt > 20~\GeVc$ and $|\eta|<$~2.4. In
addition, the track associated with the muon candidate is required to
have at least 11 hits in the silicon tracker, to be consistent with a particle
originating from the primary vertex in the event, and to
have a high-quality global fit including a minimum number of hits in
the muon detectors~\cite{wzxs}. If more than one primary vertex is
found for the same bunch crossing, only that with the highest
summed $\pt$ of the associated tracks is considered.

Electron candidates are reconstructed from clusters of energy deposits
in the ECAL, which are then matched to hits in the silicon
tracker. Seeded track trajectories are reconstructed
with a "Combinatorial track finder" algorithm, and then fitted using
a "Gaussian sum filter" algorithm, which takes into account bremsstrahlung emission
as the electron traverses tracker material~\cite{gsfFit,egmpas}. Electron
candidates are required to have $\pt > 20~\GeVc$ and
$|\eta|<$~2.5. The electron candidate track is also required to be
consistent with a particle originating from the primary vertex in the event.
Electron identification criteria based on shower shape and
track-cluster matching are applied to the reconstructed
candidates. The criteria were optimized in the context of inclusive $\W$ and
$\Z$ cross section measurements~\cite{wzxs} and are
designed to maximally reject misidentified electrons from QCD multijet
production and nonisolated electrons from heavy-quark decays, while
maintaining at least 80\% efficiency for electrons from the decay
of \W\ or \Z\ bosons. Electrons originating from photon conversions
are suppressed by looking for a partner track and requiring no missing
hits in the pixel detector for a track fit~\cite{egmpas}.

Charged leptons from $\rm W$ boson decays are expected to be isolated
from any other activity in the event. For each lepton candidate, a cone of
radius $\Delta R\equiv\sqrt{\Delta \eta^2 + \Delta \phi^2} <$~0.3 is
constructed around the track direction at the event vertex. The activity
around the lepton is determined from the scalar sum of the transverse energies
of all tracks and all deposits in the ECAL and HCAL contained in the cone,
with the exception of the lepton contributions.
If this sum exceeds 15 (10)\% of the muon $\pt$ (electron $E_{\rm{T}}$),
the candidate is not selected.

Neutrinos from ${\rm W}$ boson decays escape detection, resulting in an
imbalance of the energy in the projection perpendicular to the beam axis, called $\met$.
The $\met$ measured from calorimeter energy deposits is
corrected to take into account
the contribution from muons and information from individual tracks
reconstructed in the tracker to correct for the calorimeter response~\cite{metpas}.
The event selection requires $\met>$~20~$\GeV$ to suppress
the Drell--Yan background.

For the event selection also a derived quantity called {\it projected
$\met$}~\cite{CDF} is used.  With $\Delta\phi$ the azimuthal angle
between $\met$ and the closest lepton, the projected $\met$ is defined
as the component of $\met$ transverse to the lepton direction if
$\Delta\phi$ is smaller than $\pi/2$, and the full magnitude of
$\met$ otherwise. This variable helps to reject $\dytt$ background events as
well as \dyll\ events with misreconstructed $\met$ associated with lepton
misreconstruction. Events are selected with projected $\met$ above
35~$\GeV$ in the $\Ep\Em$ and $\Mp\Mm$ final states, and above 20
$\GeV$ in the $\Epm\Mmp$ final state that has lower
contamination from $\dyll$ decays. These requirements remove more
than 99\% of the Drell--Yan contribution.

To further reduce Drell--Yan background in the $\Ep\Em$ and $\Mp\Mm$ final
states a $\Z$ veto is defined, by which events with a dilepton
invariant mass within 15~$\GeVcc$ of the $\Z$ mass are discarded.
Events are also rejected with dilepton masses below 12~$\GeVcc$ to
suppress contributions from low mass resonances.

To reduce backgrounds containing top quarks, events containing jets
with $|\eta|<5.0$ and $\pt > 25~\GeVc$ are rejected. Jets are
clustered from the particles reconstructed with the particle-flow
event reconstruction~\cite{jetpas, pfpas1, pfpas2}, which combines
the information from all CMS sub-detectors. The anti-${\rm k_T}$ clustering
algorithm~\cite{antikt} with distance parameter ${\it R} = 0.5$ is used.
The jet veto is complemented by a {\it top veto} based on
soft-muon and b-jet tagging~\cite{btag1,btag2}. This veto allows further
rejection of top quark background and also provides a way of estimating the
remaining top quark background using the data.

To reduce the background from diboson processes, such as
$\WZ$ and $\ZZ$ production, any event which has an additional third lepton
passing the identification and isolation requirements is rejected.

\begin{figure}
\begin{center}
   \subfigure[]{\includegraphics[width=0.45\textwidth,angle=90]{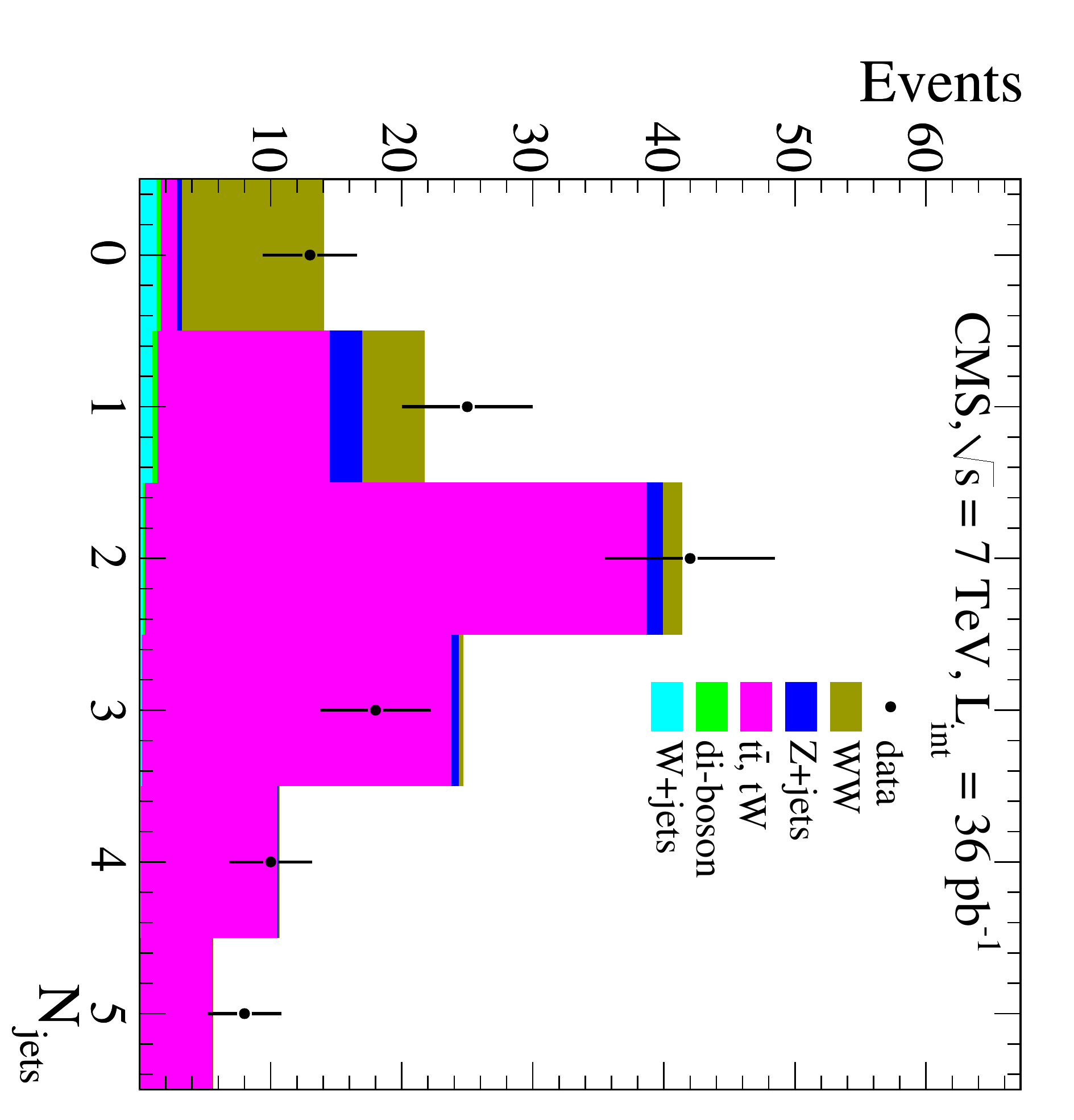}}
   \subfigure[]{\includegraphics[width=0.45\textwidth,angle=90]{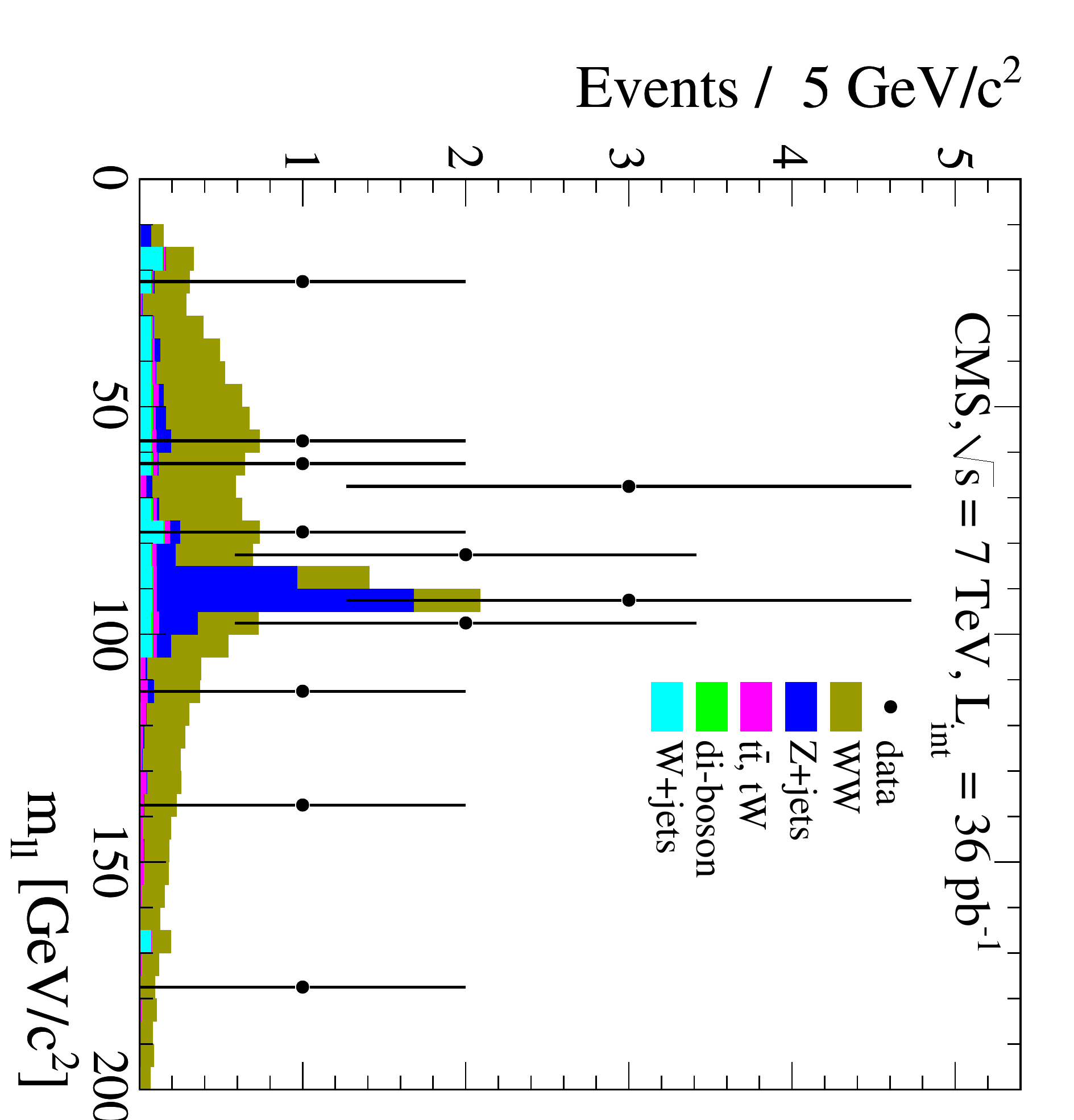}}
       \caption{
       (a) Jet multiplicity distribution after all $\WW$ selection criteria,
    except the top veto and jet veto requirements. (b) Dilepton mass distribution for the events
    passing the final selections, except the $\Z$ mass veto.}
   \label{fig:ww_njets_mll}
\end{center}
\end{figure}

Table~\ref{tab:sel_eff} shows the \WW\ efficiency, obtained from
simulation of events, where both $\W$ bosons decay leptonically. As a cross-check, kinematic distributions are
compared between data and simulation. Figure~\ref{fig:ww_njets_mll}a
shows the jet multiplicity distribution for events that pass all
selections but the jet veto and top veto.
Figure~\ref{fig:ww_njets_mll}b shows the dilepton mass distribution
for events passing the final $\WW$ event selections, except the $\Z$
mass veto.

\begin{table*}[ht]
  \begin{center}
  \caption{Selection efficiency for \wwll\ events as obtained from
  simulation. The efficiency is normalized to the total number of events
  where both $\W$ bosons decay leptonically. Selections are applied sequentially.
  The efficiencies in parenthesis are defined relative to the previous cut. }
   \label{tab:sel_eff}
  \begin{tabular} {c|lll}
\hline\hline
  Selection & e$^+$e$^-$ & e$^+\mu^-$/e$^-\mu^+$ & $\mu^+\mu^-$ \\
\hline
  lepton acceptance ($\eta$, \pt{}) & 6.9\%           & 13.4\%            & 6.6\%    \\  
  primary vertex compatibility    & 6.2\% (89.9\%)  & 12.7\% (94.9\%)   & 6.5\% (98.5\%) \\ 
  lepton isolation           & 5.2\% (83.9\%)  & 11.2\% (88.2\%)   & 6.1\% (93.8\%) \\ 
  lepton identification      & 4.1\% (78.8\%)  & ~9.6\% (85.7\%)   & 5.6\% (91.8\%) \\ 
  $\gamma$ conversion rejection& 3.9\% (95.1\%)  & ~9.4\% (97.9\%)   & 5.6\% (100.0\%) \\ 
  $\met>20~\GeV$           & 3.2\% (82.5\%)  & ~7.7\% (82.5\%)   & 4.6\% (82.4\%)  \\ 
  $\mll>12~\GeVcc$           & 3.2\% (100.0\%) & ~7.7\% (100.0\%)  & 4.6\% (100.0\%) \\ 
  Z mass veto              & 2.5\% (77.1\%)  & ~7.7\% (100.0\%)  & 3.5\% (77.2\%) \\ 
  projected $\met$      & 1.5\% (61.3\%)  & ~6.7\% (86.7\%)   & 2.2\% (63.1\%) \\ 
  jet veto            & 0.9\% (60.8\%)  & ~4.2\% (62.3\%)   & 1.4\% (61.4\%) \\ 
  extra lepton veto   & 0.9\% (100.0\%) & ~4.2\% (100.0\%)  & 1.4\% (100.0\%) \\ 
  top veto    & 0.9\% (100.0\%) & ~4.1\% (99.4\%)   & 1.4\% (100.0\%) \\ 
 \hline\hline
  \end{tabular}
  \end{center}
\end{table*}

After applying all selection requirements, 13 events are observed in
data, with 2, 10, and 1 events coming from $\Ep\Em$, $\Epm\Mmp$,
and $\Mp\Mm$ final states, respectively, in good agreement with
simulation based expectations ($13.5\pm0.3$, $2.7\pm0.1$, $2.3\pm0.2$ and
$8.5\pm0.3$ respectively).

\subsection{Background estimation}
\label{subsec:ww_bkg}
To evaluate the remaining background contributions in data, a combination of techniques
based on data and on detailed simulation studies are used.

The accurate simulation of the $\Wjets$ and ${\rm QCD}$ multijet
instrumental background suffers from  large systematic uncertainties,
which are hence estimated with a data-based approach.
A set of loosely selected lepton-like objects is defined
in a sample of events dominated by dijet production.
The probability is calculated for those objects that are misidentified as
leptons passing all lepton selection criteria. This misidentification probability,
parameterized as a function of $\pt$ and $\eta$, is then applied to a sample
of events selected using the final selection
criteria, except for one of the leptons for which the selection has been
relaxed to the looser criteria and that has failed the nominal selection.
This procedure is
validated in simulated events and applied on data.
The systematic uncertainty on this estimation is obtained by applying
the same method to another control sample with different selection criteria.
A value of 50\% is derived from a closure test, where a tight-to-loose rate derived
from QCD simulated events is applied to a $\Wjets$ simulated sample to predict the rate of
events with one real and one misidentified lepton.  The total misidentified
electron and muon background contributions are found to be
$1.2\pm0.3~(\mathrm{stat})\pm0.6~(\mathrm{syst})$ and
$0.5\pm0.3~(\mathrm{stat})\pm0.3~(\mathrm{syst})$ events, respectively.

The remaining top quark background after full event selection can be
estimated from data by counting events with either an additional soft
muon (well identified muons with $\pt>~3~\GeVc$ are considered) or at least one
b-tagged jet with \pt\ below the jet veto threshold. No events are rejected by
the top-veto in data after applying the full selection, which
is consistent with the predictions from simulation. Therefore, the top quark
background contribution is taken directly from simulation, which predicts
$0.77\pm0.05~(\rm{stat})\pm0.77~(\rm{syst})$ events, where a 100\%
systematic uncertainty is assigned as a conservative estimate of
the difference between data and simulation.

An estimate of the residual $\Z$ boson contributions in the
$\Ep\Em$ and $\Mp\Mm$ final states outside the $\Z$ mass window,
$N_{\rm out}^{\ell\ell,{\rm exp}}$,
is obtained from data in the following way. The ratio $R^{\ell\ell}_{\rm out/in}$
of the number of events outside the $\Z$ mass window to that inside is obtained from simulation.
The observed number of events inside the $\Z$ mass window in data, $N_{\rm in}^{\ell\ell}$,
from which the non-${\rm Z}$ contributions ($N_{\rm in}^{\rm non-Z}$) is subtracted,
is then scaled by $R^{\ell\ell}_{\rm out/in}$ to compute the residual $\Z$ background:
\begin{eqnarray*}
N_{\rm out}^{\ell\ell,{\rm exp}} = R^{\ell\ell}_{\rm out/in}(N_{\rm in}^{\ell\ell} - N_{\rm in}^{\rm{non-Z}}),
~\mathrm{with}~R^{\ell\ell}_{\rm out/in} = N_{\rm out}^{\ell\ell,{\rm MC}}/N_{\rm in}^{\ell\ell,{\rm MC}}.
\label{eq:dyest}
\end{eqnarray*}
The number $N_{\rm in}^{\rm{non-Z}}$ is estimated as half of the number of
$\Epm\Mmp$ events, taking into account the relative detection efficiencies
of electrons and muons. The result also includes $\WZ$ and
$\ZZ$ contributions, in which both leptons come from the same ${\rm Z}$
boson. The total Z decay contribution is estimated as
$0.2\pm0.2~(\rm{stat})\pm0.3~(\rm{syst})$ events.  The systematic
uncertainty of this method arises primarily from the dependence of
$R^{\ell\ell}_{\rm out/in}$ on the $\met$ cut.

Other backgrounds are estimated from simulation. The $\W\gamma$ production,
where the photon is misidentified as an electron, is suppressed by
the $\gamma$ conversion rejection requirements. As a
cross-check, this background was studied using the events passing all
selection requirements, except that the two leptons must have
the same charge. This sample is dominated by $\W$ + jets and $\W\gamma$ events.
Other minor backgrounds are WZ and ZZ diboson production where
the selected leptons come from different bosons, and $\dytt$ production. All
background predictions are summarized in
Table~\ref{tab:bkg_estimation}. The estimated number of remaining
background events is $3.29\pm0.45~(\rm{stat})\pm1.09~(\rm{syst})$.

\begin{table}[tpb]
  \begin{center}
  \caption[Background estimates]{Summary of background estimations
    for $\WW\to 2\ell2\nu$ at $\sqrt{s}$ = 7~$\TeV$, corresponding to an integrated luminosity of $\intlumi$.
    Statistical and systematic uncertainties are reported.
  }
  \label{tab:bkg_estimation}
     \begin{tabular}{c|c}
     \hline \hline
     Process & Events  \\
     \hline
     \wjets\ + ${\rm QCD}$ &  $1.70\pm0.40\pm 0.70$ \\
     t$\bar{\rm{t}}$ + \tw\  &   $0.77\pm0.05\pm0.77$ \\
     $\wgamma$ & $0.31\pm0.04\pm0.05$ \\
     $\Z + WZ + ZZ\to \Ep\Em/\Mp\Mm$  & $0.20\pm0.20\pm0.30$ \\
     $\WZ + ZZ$, leptons not from the same boson  &  $0.22\pm0.01\pm0.04$  \\
     $\dytt$ &  $0.09\pm0.05\pm0.09$  \\
     Total & $3.29\pm0.45\pm1.09$ \\
     \hline \hline
     \end{tabular}
  \end{center}
\end{table}

\subsection{Efficiencies and systematic uncertainties}
\label{subsec:ww_syst}
The $\WW$ signal efficiency is estimated using the simulation, corrected by
data-to-simulation {\it scale factors}. For electron and muon reconstruction and
identification, a tag-and-probe method~\cite{wzxs} is applied to leptons from
$\dyll$ events in the $\Z$ resonance region, both in data and simulation. The
scale factors are found to be
consistent with unity for muons. For electrons, they are found to be
$(96.9\pm1.9)$\% and $(99.2\pm2.6)$\% in the barrel ($|\eta|<1.479$) and
end-cap ($|\eta| \geq 1.479$) regions, respectively.
For estimating the effect of the jet veto efficiency on the $\WW$ signal,
events in the Z resonance region are used according to the following
relation: $\epsilon_{\WW}^{\rm data} = \epsilon_{\WW}^{\rm MC} \times \epsilon_{\Z}^{\rm
data}/\epsilon_{\Z}^{\rm MC}$. The scale factor
is found to be consistent with unity. The uncertainty is factorized into the
uncertainty on the $\Z$ efficiency in data ($\epsilon_{\Z}^{\rm data}$) and
the uncertainty on the ratio of the $\WW$ efficiency to the $\Z$ efficiency in simulation
($\epsilon_{\WW}^{\rm MC}/\epsilon_{\Z}^{\rm MC}$). The uncertainty
on the former, which is statistically dominated, is 0.3\%.
Theoretical uncertainties due to higher-order corrections contribute
most to the $\WW$/$\Z$ efficiency ratio uncertainty, which is
estimated to be 5.5\% for $\WW$ production from the uncertainties on
the jet kinematics for $\WW$ and $\Z$ events from different NLO Monte
Carlo generators.

The acceptance uncertainties due to PDF
choice range from 2\% to 4\% for the different
processes~\cite{PDF4LHC,Huston:2010zz}. The uncertainties from lepton
identification and trigger requirements range from 1\% to 4\%. The
effect on the signal efficiency from multiple collisions within a bunch crossing
is $0.5\%$, as evaluated by reweighting the number of
reconstructed primary vertices in simulation to match the distribution
found in data.  The uncertainty from the luminosity measurement is
11\%~\cite{lumiPAS}. Overall the uncertainty is estimated to be 7\%
on the $\WW$ selection efficiency, coming mainly from the theoretical
uncertainty in the jet veto efficiency determination.
The uncertainty on the background estimations in the $\WW$ signal region,
reported in Table~\ref{tab:bkg_estimation}, is about 37\%,
dominated by statistical uncertainties in the data control regions.

\subsection{$\WW$ cross section measurement}
\label{subsec:ww_results}
The $\WW$ yield is calculated from the number of events in the signal
region, after subtracting the expected contributions
of the various SM background processes. From this yield and  the W$ \to \ell \nu$
branching fraction~\cite{pdg}, the $\WW$ production
cross section in pp collisions at $\sqrt{s} = $ 7~TeV is found to be
\begin{displaymath}
\sigma_{\WW}  = 41.1 \pm 15.3 \,(\rm{stat}) \pm 5.8 \,(\rm{syst}) \pm 4.5 \,(\rm{lumi})\,\rm{pb}.
\end{displaymath}
This measurement is consistent with the SM expectation of $43.0 \pm
2.0$ pb at NLO~\cite{MCFM}.

The WW to W cross section ratio is also computed. In this ratio, the luminosity
uncertainty cancels out, and uncertainties for the signal
efficiency and background contamination can be considered mostly uncorrelated,
since the correlated factors form a very small fraction of the
overall uncertainty. The W$\to \ell \nu$ cross section is taken
from Ref.~\cite{wzxs} to obtain the following cross section
ratio:

\begin{equation*}
\frac{\sigma_{WW}}{\sigma_{W}} = (4.46 \pm 1.66 \pm 0.64) \cdot 10^{-4},
\end{equation*}

in agreement with the expected theoretical ratio
$(4.45 \pm 0.30) \cdot 10^{-4}$~\cite{Melnikov:2006kv, Melnikov:2006di,MCFM}.

\section{Limits on WW$\gamma$ and WWZ Anomalous Triple Gauge Couplings}
\label{sec:aTGCs}
A measurement of triple gauge couplings is performed and limits on
anomalous couplings are set, using the
effective Lagrangian approach with the HISZ
parametrization~\cite{HISZ} without form factors. Three parameters,
$\lambda_{\Z}$, $\kappa_\gamma$, and $g_1^Z$,
are used to describe all operators which are Lorentz and
$SU(2)_L$ $\otimes$ $U(1)_Y$ invariant and conserve $C$ and $P$ separately. In the
SM, $\lambda_{\Z} = 0$ and $\kappa_\gamma = g_1^Z = 1$. In this paper,
$\Delta\kappa_\gamma$ and $\Delta g_1^{\Z}$ are used to denote the
deviation of the $\kappa_\gamma$ and $g_1^Z$ parameters with
respect from the SM values. Two different measurements of the
anomalous couplings are performed. Both
use the leading lepton $\pt$ distribution. The first measurement uses
a binned fit, while the second uses an unbinned fit to data.
The uncertainties on the quoted luminosity, signal selection, and
background fraction are assumed to be Gaussian, and are reflected in the
likelihood function used to determine the limits in the form of nuisance
parameters with Gaussian constraints.

\begin{figure}[htb]
  \begin{center} \includegraphics[width=0.60\textwidth]{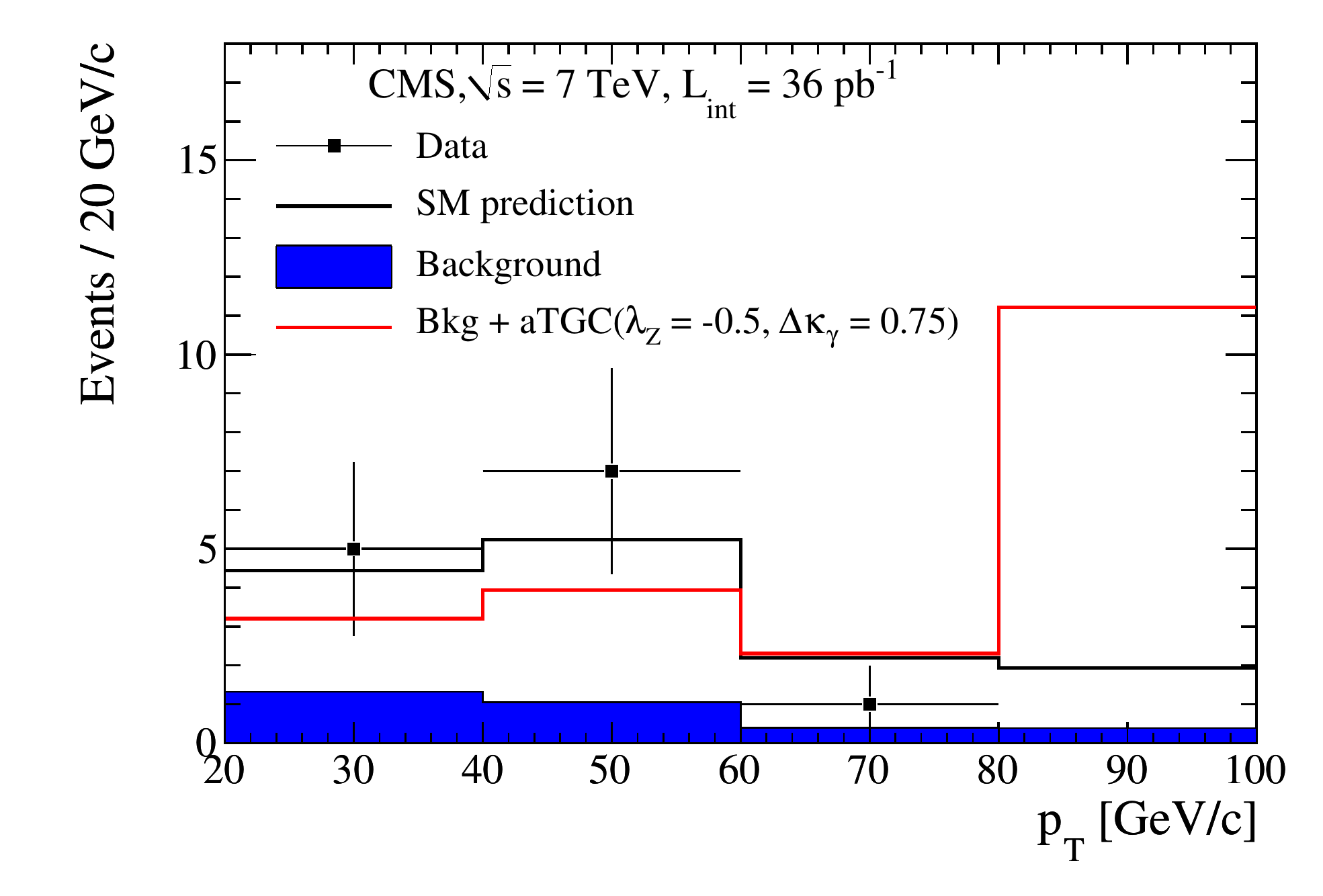}
  \caption[Leading lepton \pt{}]{Leading lepton \pt\ distribution in data overlaid
    with predictions from the SM simulation, background only simulation ({\it Bkg} in
    the figure) and the simulation with large anomalous
    couplings ({\it aTGC} in the figure).}  \label{fig:atgc_pt} \end{center}
\end{figure}

\begin{figure}[tp]
  \centerline{
    \subfigure[]{\includegraphics[width=0.49\textwidth]{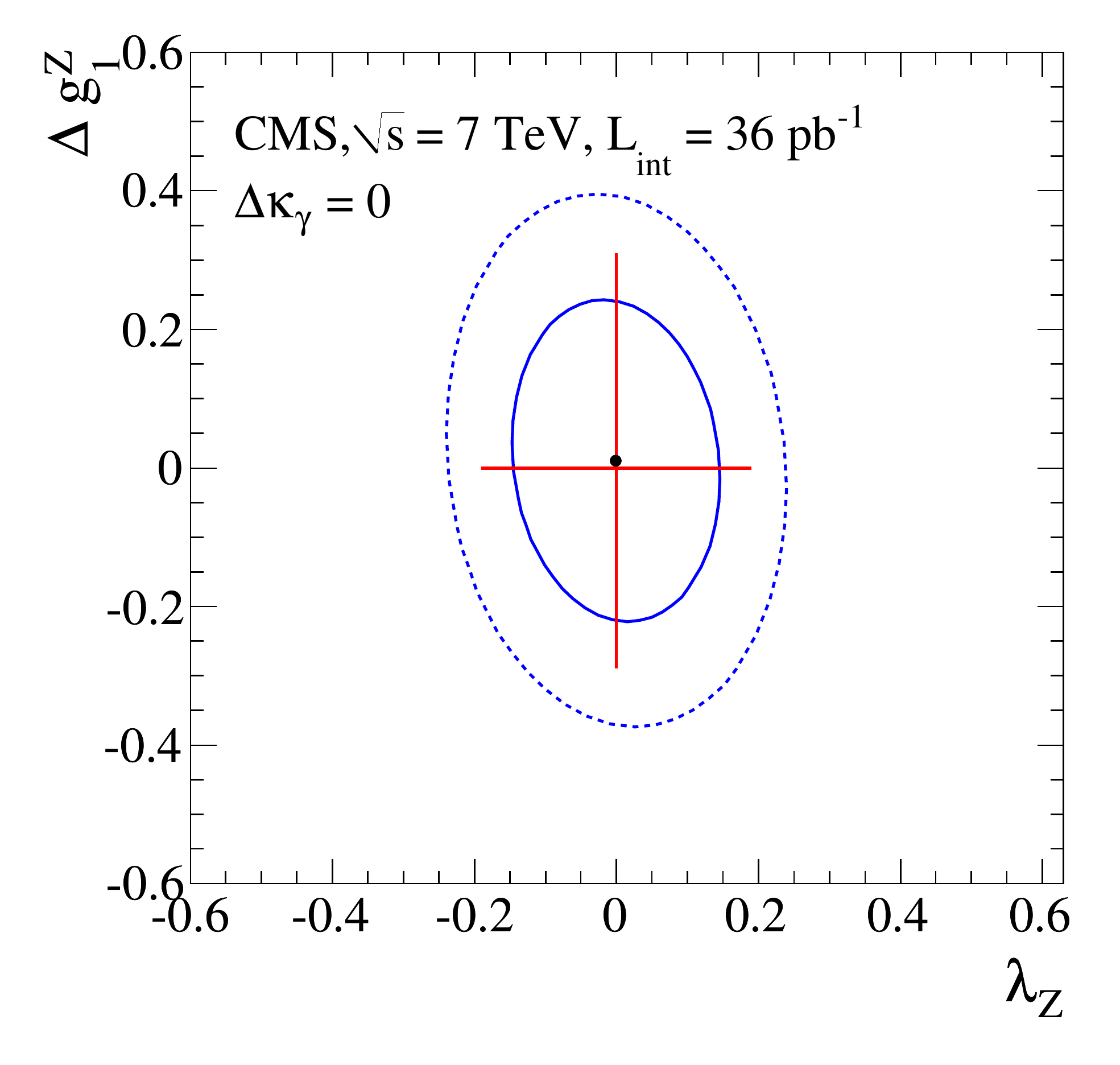}}
    \subfigure[]{\includegraphics[width=0.49\textwidth]{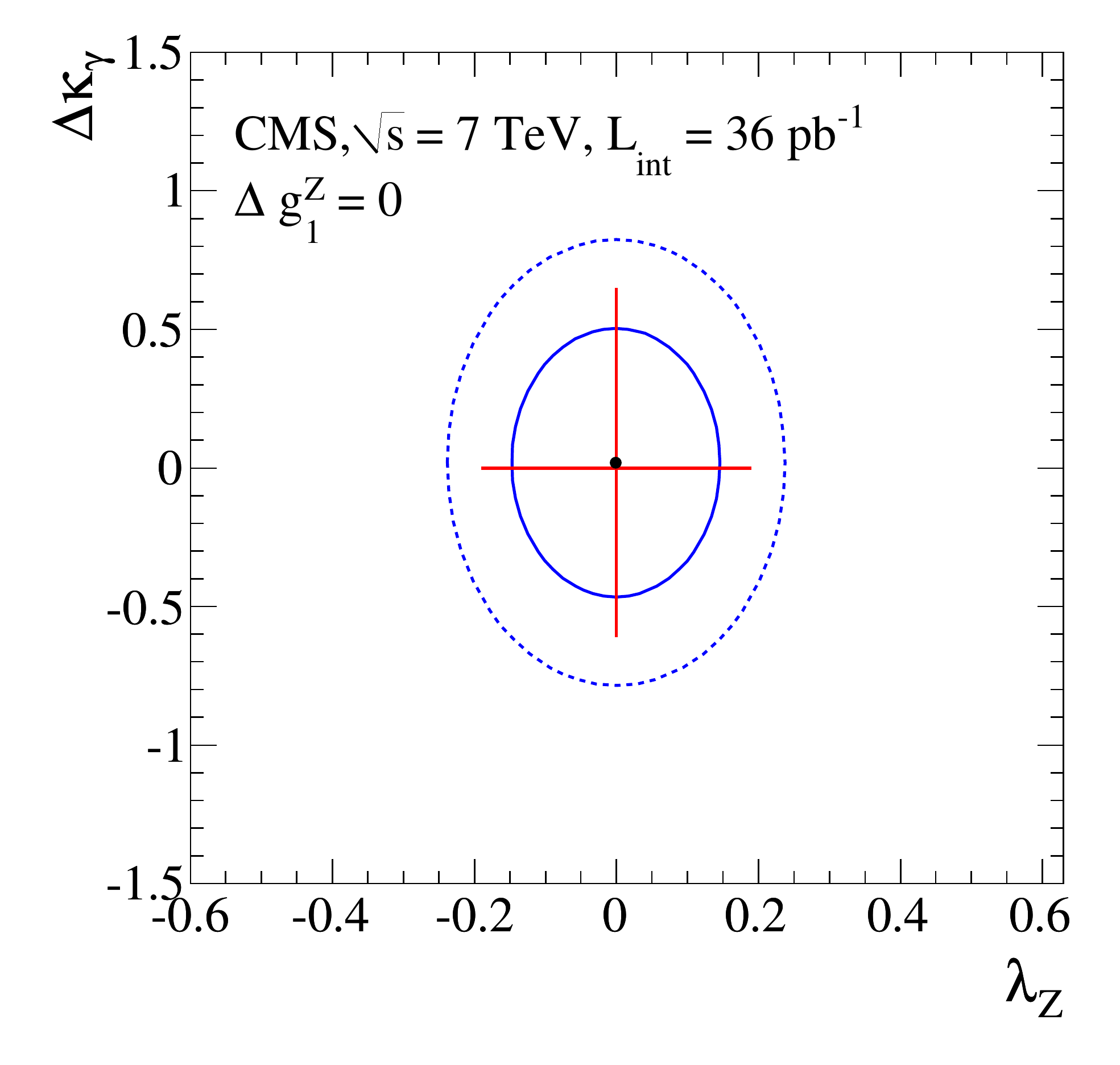}}
  }
  \caption{68\% (solid blue lines) and 95\% C.L. (dotted blue lines) as well
    as the central value (point) and one-dimensional 95\% C.L. limits (red lines)
    using unbinned fits,
    for (a) $\Delta \kappa_\gamma = 0$ and (b) $\Delta g_1^{\Z} = 0$.
  }
  \label{fig:contours}
\end{figure}

Figure~\ref{fig:atgc_pt} shows the leading lepton \pt\ distributions
in data and the predictions for the SM $\WW$ signal and background
processes, and for a set of large anomalous couplings.
Table~\ref{tab:atgc_results} presents the 95\% C.L. limits
on one-dimensional fit results for anomalous TGC that correspond to the change
in the log-likelihood of 1.92. Both methods give similar results,
consistent with the SM. The limits are comparable to the current
Tevatron results~\cite{CDF, D0}.  In Fig.~\ref{fig:contours} the
contour plots of the 68\% and 95\%~C.L. for the $\Delta\kappa_\gamma =
0$ and $\Delta g_1^{\Z} = 0$ scenarios are displayed. The contours
correspond to the change in the log-likelihood of 1.15 and 2.99
respectively.

\begin{table}[htb]
\begin{center}
\caption[aTGC limits]{ 95\% C.L. limits on one-dimensional fit results for anomalous TGC.}
\label{tab:atgc_results}
\begin{tabular}{c|c|c|c} \hline \hline
           &  $\lambda_{\Z}$ & \rule{0mm}{4.2mm}$\Delta g_1^{\Z}$ & $\Delta\kappa_\gamma$\\
  \hline
  Unbinned fit & $[-0.19, 0.19]$ & $[-0.29, 0.31]$ & $[-0.61, 0.65]$ \\
  Binned fit & $[-0.23, 0.23]$ & $[-0.33, 0.40]$ & $[-0.75, 0.72]$ \\
\hline
\hline
\end{tabular}
\end{center}
\end{table}

\section{Search for Higgs Bosons in the $\WW$ Decay Mode}
\label{sec:hww}
The preselection for the Higgs boson search in the $\WW$ decay mode is
identical to the $\WW$ selection described in
Section~\ref{subsec:ww_evtsel}. To enhance the sensitivity to the
Higgs boson signal, two different analyses are performed. The first analysis
is a cut-based approach where further requirements on a few
observables are applied, while the second analysis makes use of
multivariate techniques. Both of them cover a large Higgs boson mass
($m_{\rm{H}}$) range, and each is separately optimized for different
$m_{\rm{H}}$ hypotheses. The first method is the simplest approach to
be performed on the limited recorded data sample. The second one is
more powerful, since it exploits the information present in the
correlation among the variables.

\subsection{Search strategy}
\label{subsec:hww_strategy}

\begin{figure}[ht]
\begin{center}
  \includegraphics[width=0.49\textwidth,angle=90]{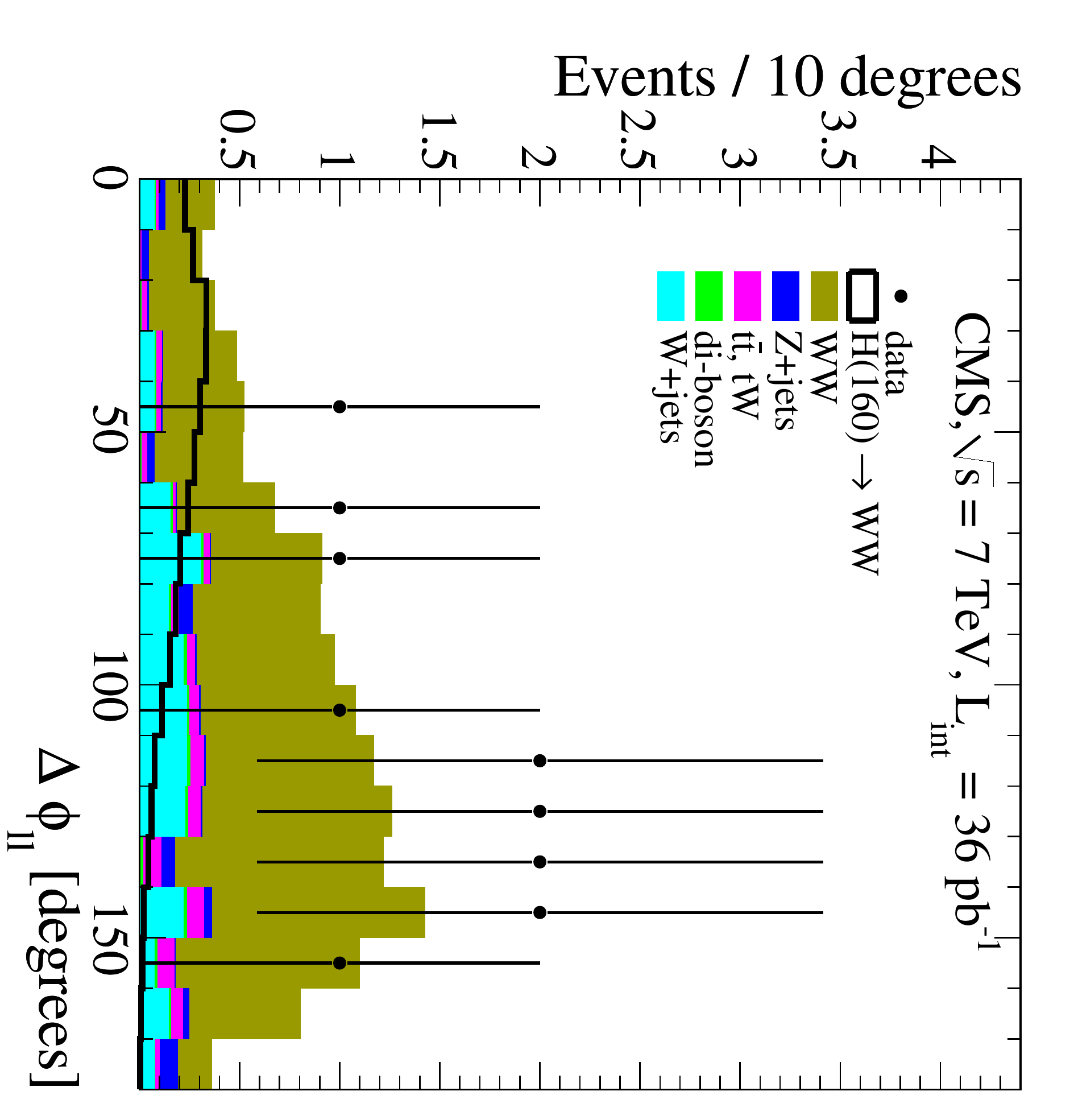}
       \caption{Azimuthal angular separation between the two selected leptons after $\WW$
       selection, for $\mHi=160~\GeVcc$ SM Higgs signal and for backgrounds. The area marked as
       $\WW$ corresponds to the nonresonant contribution.}
   \label{fig:deltaphi}
\end{center}
\end{figure}

In the cut-based approach, the extra selections are based on the transverse momenta
of the harder (\ptlmax ) and the softer (\ptlmin) leptons, the dilepton mass $\mll$,
and the azimuthal angle difference $\delphill$ between the two selected leptons.
Among these variables, $\delphill$ provides the best discriminating
power between the Higgs boson signal and the majority of the backgrounds in
the low mass range~\cite{HWW}. Leptons originating from
$\hww$ decays tend to have a relatively small opening angle, while those from
backgrounds are preferentially emitted back-to-back.
Figure~\ref{fig:deltaphi} shows the $\delphill$ distribution,
after applying the $\WW$ selections, for a SM Higgs boson signal
with $\mHi=160~\GeVcc$, and for backgrounds.

Because the kinematic properties of the Higgs boson decay depend on its mass,
the selection criteria were optimized for each assumed mass value.
The requirements are summarized in
Table~\ref{tab:cuts_analysis}. The numbers of events observed in $\intlumi$
of data, with the signal and background predictions are listed
in Table~\ref{tab:hwwselection}.

\begin{table*}[ht]
  \begin{center}
  \caption{Values of the selection requirements for several \mHi\ mass hypotheses. }
  \label{tab:cuts_analysis}
    \begin{tabular}{c|c|c|c|c}
    \hline\hline
\mHi  & $\ptlmax$ (\GeVc) & $\ptlmin$ (\GeVc) & $\mll$ (\GeVcc) & $\delphill$ (degree) \\
(\GeVcc)     &   $>$               &   $>$               &   $<$             &  $<$              \\  \hline
    130 & 25  &  20 & 45  & 60  \\
    160 & 30  &  25 & 50  & 60  \\
    200 & 40  &  25 & 90  & 100 \\
    210 & 44  &  25 & 110 & 110 \\
    400 & 90  &  25 & 300 & 175 \\
      \hline \hline
    \end{tabular}
  \end{center}
\end{table*}

\begin{table*}[!ht]
  \begin{center}
  \caption{Numbers of events observed in $\intlumi$ of data, with the signal and background predictions after $\hww$ selections
in both cut-based and multivariate approaches. Only statistical uncertainties from the simulations are included.}
   \label{tab:hwwselection}
 {\scriptsize
  \begin{tabular} {c|c|c|c|c|c|c|c}
\hline\hline
$m_H$  & \multirow{2}{*} {data}  & SM &  SM with 4th gen. & \multirow{2}{*}{all bkg.}
& \multirow{2}{*}{qq$\to \WW$} & \multirow{2}{*} {gg$\to \WW$}  & all non-\\
(\GeVcc) & & $\hww$ & $\hww$ & & & &  $\WW$ \\
\hline
\multicolumn{8}{c}{cut-based approach} \\
  \hline
130 & 1 & 0.30 $\pm$ 0.01 &  1.73 $\pm$ 0.04 & 1.67 $\pm$  0.10 & 1.12 $\pm$  0.01 & 0.10 $\pm$  0.01 & 0.45 $\pm$  0.10 \\
160 & 0 & 1.23 $\pm$ 0.02 & 10.35 $\pm$ 0.16 & 0.91 $\pm$  0.05 & 0.63 $\pm$  0.01 & 0.07 $\pm$  0.01 & 0.21 $\pm$  0.05 \\
200 & 0 & 0.47 $\pm$ 0.01 &  3.94 $\pm$ 0.07 & 1.47 $\pm$  0.09 & 1.13 $\pm$  0.01 & 0.12 $\pm$  0.01 & 0.23 $\pm$  0.09 \\
210 & 0 & 0.34 $\pm$ 0.01 &  2.81 $\pm$ 0.07 & 1.49 $\pm$  0.05 & 1.09 $\pm$  0.01 & 0.10 $\pm$  0.01 & 0.30 $\pm$  0.05 \\
400 & 0 & 0.19 $\pm$ 0.01 &  0.84 $\pm$ 0.01 & 1.06 $\pm$  0.03 & 0.79 $\pm$  0.01 & 0.04 $\pm$  0.01 & 0.23 $\pm$  0.03 \\
  \hline
\multicolumn{8}{c}{multivariate approach} \\
  \hline
130 & 1 & 0.34 $\pm$  0.01 &  1.98 $\pm$  0.04 & 1.32 $\pm$ 0.18 &  0.75 $\pm$  0.01 &  0.04 $\pm$  0.00  & 0.53 $\pm$  0.18\\
160 & 0 & 1.47 $\pm$  0.02 & 12.31 $\pm$  0.17 & 0.92 $\pm$ 0.10 &  0.63 $\pm$  0.01 &  0.06 $\pm$  0.00  & 0.22 $\pm$  0.10\\
200 & 0 & 0.57 $\pm$  0.01 &  4.76 $\pm$  0.07 & 1.47 $\pm$ 0.07 &  1.07 $\pm$  0.01 &  0.13 $\pm$  0.00  & 0.27 $\pm$  0.07\\
210 & 0 & 0.42 $\pm$  0.01 &  3.47 $\pm$  0.07 & 1.44 $\pm$ 0.07 &  1.03 $\pm$  0.01 &  0.12 $\pm$  0.00  & 0.29 $\pm$  0.07\\
400 & 0 & 0.20 $\pm$  0.01 &  0.90 $\pm$  0.01 & 1.09 $\pm$ 0.07 &  0.75 $\pm$  0.01 &  0.04 $\pm$  0.00  & 0.30 $\pm$  0.07\\
 \hline
 \hline
  \end{tabular}
  }
  \end{center}
\end{table*}

In the multivariate approach a boosted decision tree (BDT) technique~\cite{tmva} is used
for each Higgs boson mass hypothesis.
In addition to the $\WW$ selection requirements, a loose cut on the
maximum value of $\mll$ is applied to enhance the signal-to-background ratio.
The multivariate technique uses the following additional variables compared to the cut-based analysis:
$\Delta R_{\Lep\Lep}\equiv\sqrt{\deletall^2 + \delphill^2}$
between the leptons, $\deletall$ being the $\eta$ difference between the leptons,
which has similar properties as $\delphill$; the angles in the transverse plane between
$\met$ and each lepton, which discriminates against events with
no real $\met$; the projected $\met$;
the transverse mass of both lepton-$\met$ pairs; and finally lepton flavours.

\begin{figure}
\begin{center}
   \subfigure[]{\includegraphics[width=0.47\textwidth,angle=90]{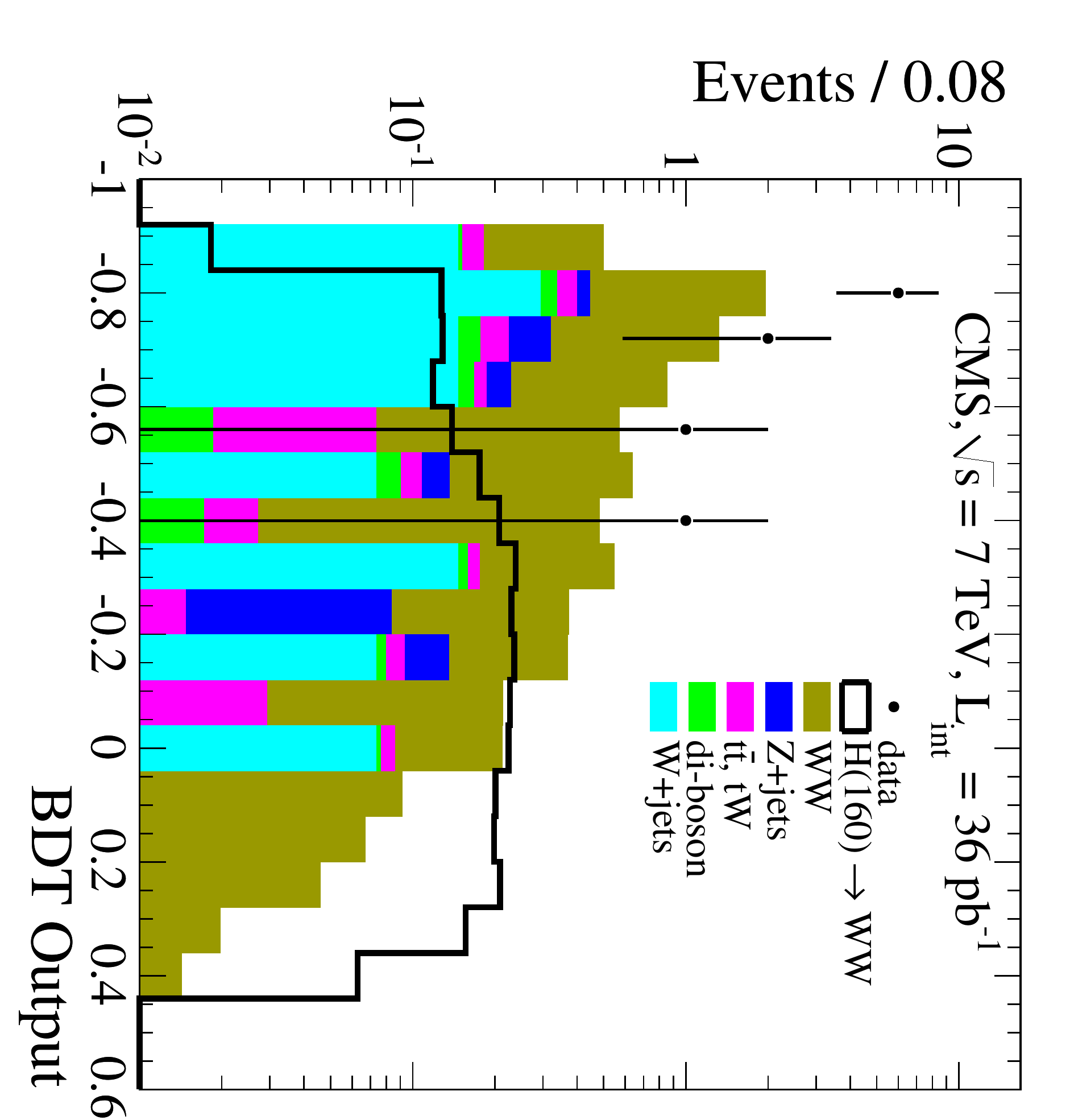}}
   \subfigure[]{\includegraphics[width=0.47\textwidth,angle=90]{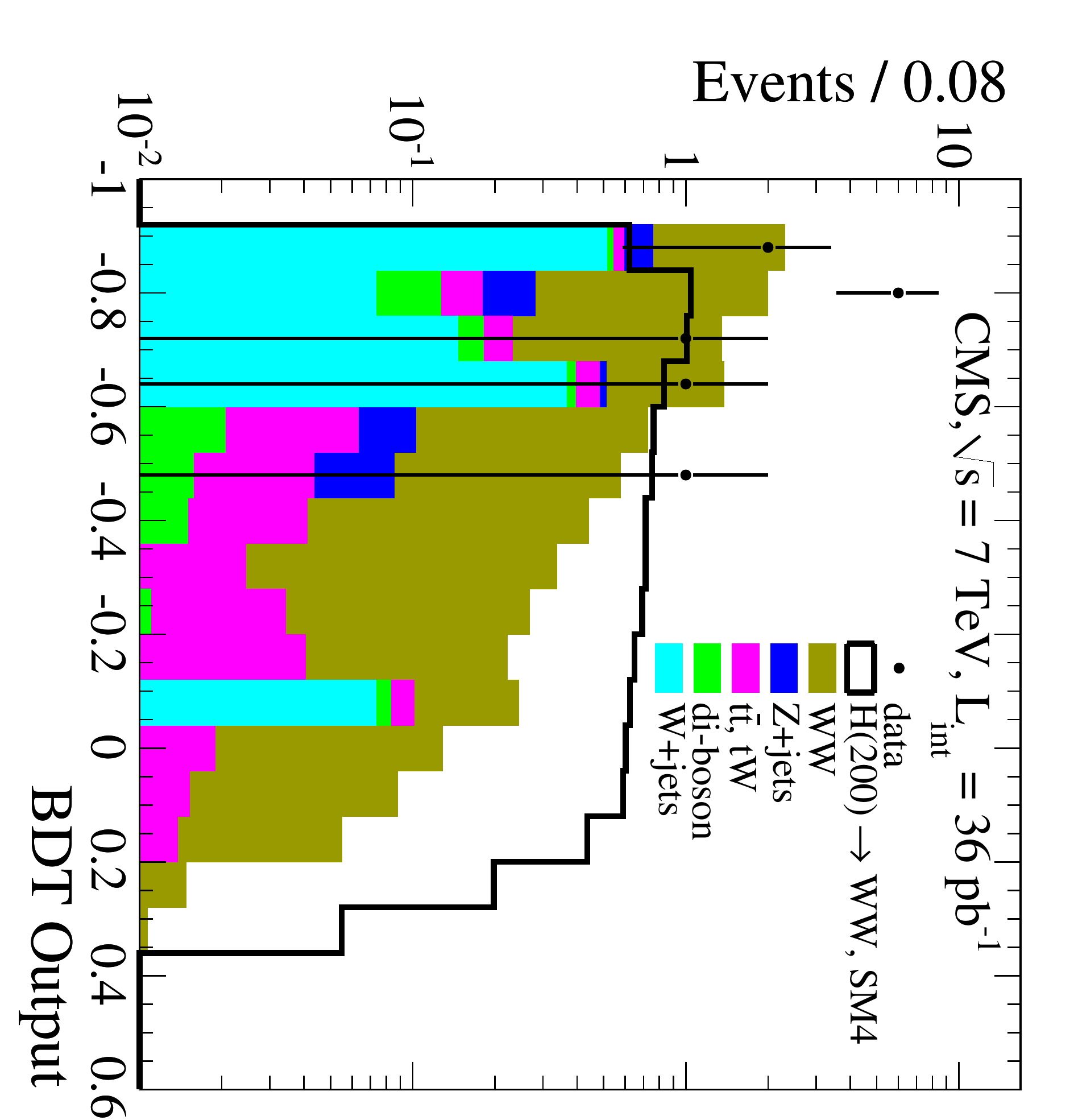}}
       \caption{BDT outputs for Higgs boson signal and for backgrounds, for (a) $\mHi=160~\GeVcc$ and
       (b) $\mHi=200~\GeVcc$.
       The Higgs boson event yield is normalized (a) to the SM expectation, and (b) to the fourth family scenario.
       The area marked as $\WW$ corresponds to the nonresonant contribution.}
   \label{fig:tmva_outputs_analysis}
\end{center}
\end{figure}

The BDT outputs for $\mHi=160~\GeVcc$ and $\mHi=200~\GeVcc$ are shown in Fig.~\ref{fig:tmva_outputs_analysis}.
The Higgs boson event yield is normalized to the SM expectation in Fig.~\ref{fig:tmva_outputs_analysis}a,
while in Fig.~\ref{fig:tmva_outputs_analysis}b the normalization is to the fourth family scenario.
The cut on the BDT output is chosen to have similar levels of background
as the cut-based analysis.
Given the better discriminating power of the BDT analysis, the
corresponding signal yields for each Higgs boson mass are about 15\% higher
than those obtained using the cut-based selection.
The numbers of events observed in $\intlumi$ of data and the signal and background predictions
are compared in Table~\ref{tab:hwwselection}.

\subsection{Background estimation}
\label{subsec:hww_bkg}

The nonresonant $\WW$ contribution in the $\hww$ signal region is estimated
from data using the dilepton mass distribution. For a given Higgs boson mass, the
region with a small contribution from Higgs boson decays is selected and simulation
is used to extrapolate this background into the signal region.
For low Higgs boson mass values ($m_{\rm{H}} < 200~\GeVcc$) events with
$m_{\ell\ell} > 100~\GeVcc$ are used, while for $m_{\rm{H}} > 200~\GeVcc$ events with
$m_{\ell\ell} < 100~\GeVcc$ are used.
The statistical uncertainty on the estimate of the nonresonant $\WW$ background with the
current data sample is approximately 50\%.

The non-$\WW$ backgrounds are estimated in the same way as in the $\WW$ production
cross section measurement described in Section~\ref{subsec:ww_bkg}.
The $\Wjets$, ${\rm QCD}$ and Drell--Yan $\dyll$ backgrounds are
estimated using data. For each studied Higgs mass region,
the \WW\ background estimated in the complementary mass region is then extrapolated
into the studied region, taking into account the effects of the selection criteria,
as determined from simulation.

In addition, for the present measurement other Higgs boson production mechanisms are considered
as backgrounds: a Higgs boson in the final state accompanied by a $\W$ or $\Z$ boson or by
a pair of top quarks, and the vector boson fusion process. These processes are
heavily suppressed by the jet and additional lepton veto requirements, and the corresponding yield
amounts to 1--2\% of the gluon fusion process.

\subsection{Systematic uncertainties}
\label{subsec:hww_syst}
Systematic uncertainties related to acceptance and efficiencies for $\hww$
are estimated in a similar way as described in Section~\ref{subsec:ww_syst}.

Simulated events are used to predict the $\hww$ signal efficiency, and
$\zll$ events are used to study the data-to-simulation efficiency scale factors of
the lepton selection and jet veto requirement. Due to details in the
implementation of the {\sc POWHEG} calculation~\cite{powheg_h}, the resulting Higgs boson
$\pt$ spectrum is harder than the most precise spectrum
calculated~\cite{NNLO_h} to NNLO with
resummation to next-to-next-leading-log (NNLL) order. Therefore, the Higgs boson $\pt$
distribution is reweighted in {\sc POWHEG} to match the NNLO+NNLL prediction.
The signal efficiency, estimated after this reweighting, is 14\% larger than that
from uncorrected {\sc POWHEG} calculations, and it is independent of the Higgs boson mass.

The overall uncertainty on the $\hww$ signal yield is estimated to be of about 14\%, where
the uncertainty on the jet veto efficiency and the luminosity determination are the
main contributions. The uncertainties on the background estimations in the $\hww$ signal
regions are about 40\%, dominated by statistical uncertainties in the data control regions.

\subsection{Results}
\label{sec:hwwlimits}

Upper limits are derived on the product of the gluon fusion Higgs
boson production cross section by the H$\to \WW$ branching fraction,
$\sigma_{\rm{H}}~\cdot~$BR(H $\to \WW \to 2\ell2\nu)$. Two different statistical methods are
used, both using the same likelihood function from the expected number of observed events
modeled as a Poisson random variable whose mean value is the sum of the
contributions from signal and background processes.
The first method is based on Bayesian inference~\cite{bayesian}, while the
second method, known as $CL_{s}$, is based on the hybrid
Frequentist-Bayesian approach~\cite{Cousins}. Both methods account for
systematic uncertainties. Although not identical,
the upper limits obtained from both methods are very similar. Results are reported
in the following using only the Bayesian approach, with a flat signal {\it prior}.

\begin{figure}[htb]
  \begin{center}
   \subfigure[]{\includegraphics[width=0.49\textwidth]{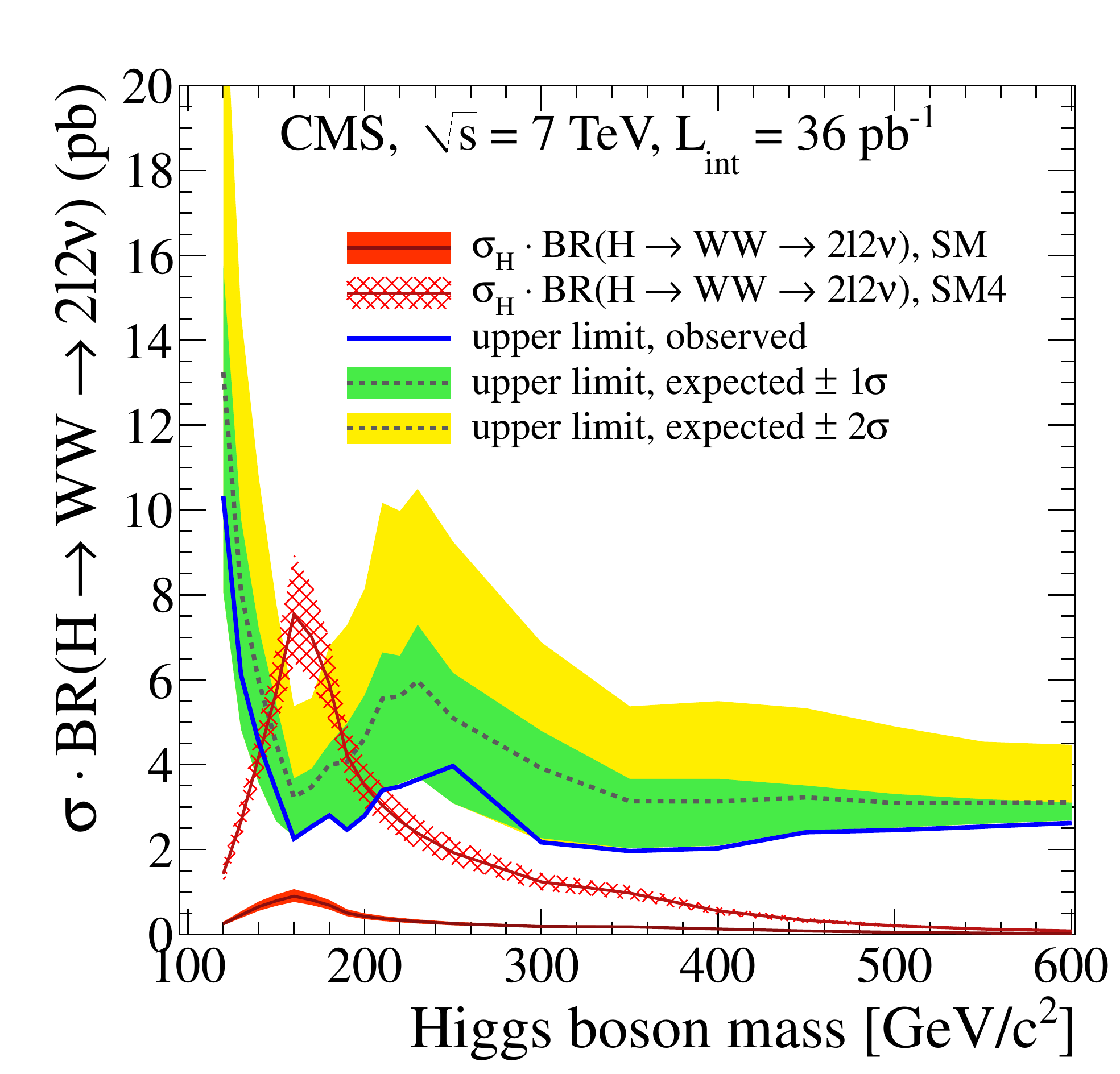}}
   \subfigure[]{\includegraphics[width=0.49\textwidth]{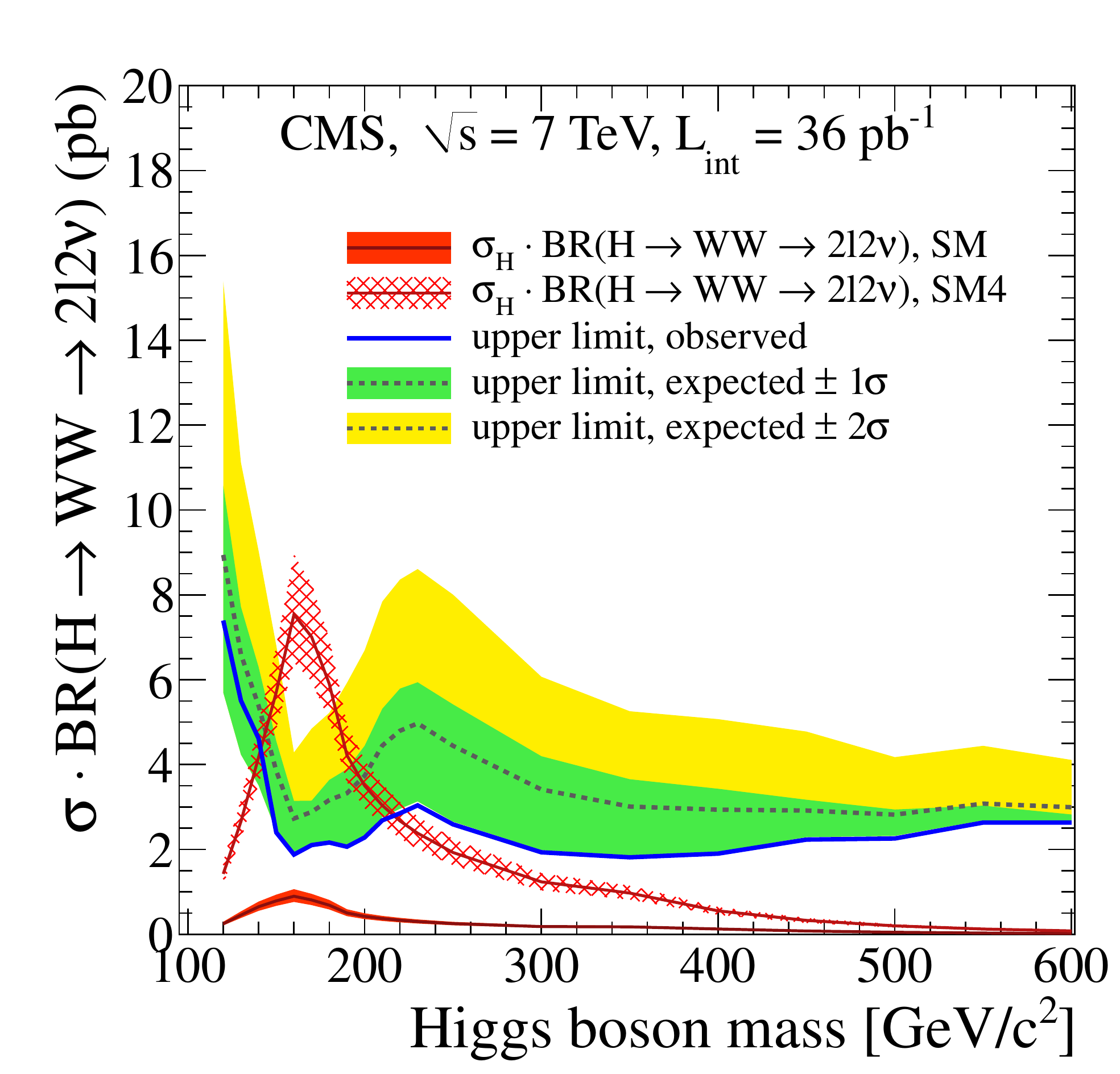}}
    \caption{95\% mean expected and observed C.L. upper
       limits on the cross section $\sigma_{\rm{H}}~\cdot~$BR(H $\to \WW \to 2\ell2\nu)$ for masses in the range
       120-600~$\GeVcc$ using (a) cut-based and (b) multivariate BDT event selections.
       Results are obtained using a Bayesian approach. The expected
       cross sections for the SM and for the SM with a fourth-fermion family cases (SM4) are also presented.
       The dash line indicates the mean of the expected results.}
    \label{fig:xsLimCuts}
  \end{center}
\end{figure}

The 95\% observed and mean expected C.L. upper limits on
$\sigma_{\rm{H}}~\cdot~$BR(H $\to \WW \to 2\ell2\nu)$ are given in
Table~\ref{tab:hwwresults} for several masses, and shown in
Fig.~\ref{fig:xsLimCuts} for Higgs boson masses in the range
120-600~$\GeVcc$. Results are reported
for both the cut-based and the BDT event selections, along with the expected
cross sections for the SM case and for the fourth-fermion family case. The
bands represent the $1 \sigma$ and $2 \sigma$ probability intervals around the
expected limit. The {\it a posteriori} probability intervals on the cross section are
constrained by the {\it a priori} minimal assumption that the signal and background
cross sections are positive definite. The expected background yield is small,
hence the $1 \sigma$ range of expected outcomes includes pseudo-experiments
with zero observed events. The lower edge of the $1 \sigma$ band therefore corresponds
already to the most stringent limit on the signal cross section, and fluctuations
below that value are not possible.

The $\sigma_{\rm{H}}~\cdot~$BR(H $\to \WW \to 2\ell2\nu)$  upper limits are
about three times larger than the SM expectation for $\mHi=160~\GeVcc$. When
compared with recent theoretical calculations performed in the context
of a SM extension by a sequential fourth family of fermions with
very high masses~\cite{4thfexp, 4thftheory},
the results of BDT analyses exclude at 95\% C.L. a Higgs boson with mass in the range
from 144 to 207 $\GeVcc$. Similar results are achieved using the
cut-based approach. The drop in the expected cross section upper limit
in the Higgs boson mass region between 200 and 250 $\GeVcc$ is due to the
lower signal efficiency while the background expectation remains at similar
levels.

\begin{table*}[!ht]
  \begin{center}
  \caption{95\% observed and mean expected C.L. upper
       limits on the cross section $\sigma_{\rm{H}} \cdot $BR(H $\to \WW \to 2\ell2\nu)$ for four Higgs masses.
       The results of the cut-based and the multivariate-based event selections are
       obtained using a Bayesian approach. The expected production
       cross sections for a SM Higgs boson~\cite{LHCHiggsCrossSectionWorkingGroup:2011ti} and
       for the scenario with an additional fourth
       family of fermions are also included.}
   \label{tab:hwwresults}
 {\footnotesize
  \begin{tabular} {c|c|c|c|c|c|c}
  \hline\hline
  \mHi & $\sigma \cdot BR$ & $\sigma \cdot BR$ & lim. obs. & lim. exp. & lim. obs. &  lim. exp. \\
(\GeVcc) & SM (pb) & 4th gen. (pb) & cut-based (pb) & cut-based (pb) & BDT-based (pb) & BDT-based (pb) \\
  \hline
130 & 0.45 & 2.66 & 6.30 & 8.07 & 5.66 & 6.57 \\
160 & 0.90 & 7.54 & 2.29 & 3.22 & 1.93 & 2.72 \\
200 & 0.42 & 3.50 & 2.80 & 4.59 & 2.32 & 3.72 \\
210 & 0.37 & 3.04 & 3.41 & 5.53 & 2.76 & 4.43 \\
400 & 0.13 & 0.55 & 2.08 & 3.12 & 1.94 & 2.93 \\
 \hline\hline
  \end{tabular}

  }
  \end{center}
\end{table*}

 \section{Summary}
 \label{sec:summary}
This paper reports the first measurement of the $\WW$ cross section and a
search for the Higgs boson decaying to $\WW$ in pp collisions at $\sqrt{s} = $ 7~TeV,
in a data sample corresponding to an integrated luminosity of $\intlumi$.
Thirteen $\WW$ candidate events, where both $\W$ bosons decay leptonically, have been observed
in the signal region, with an estimated background contribution of
$3.29\pm0.45~(\rm{stat})\pm1.09~(\rm{syst})$. The $\WW$ cross section has been measured to be
$41.1 \pm 15.3 \,(\rm{stat}) \pm 5.8 \,(\rm{syst}) \pm 4.5 \,(\rm{lumi})\,\rm{pb}$,
consistent with the SM prediction.

The $\WW$ events have been used to measure the $\W\W\gamma$ and $\W\W\Z$ triple gauge
couplings. The results, which are in agreement with the SM
predictions, are consistent with the precise measurements made at LEP
and comparable in sensitivity with the current Tevatron results.

Limits on the Higgs boson production cross section have been derived. No
excess above the SM expectations was found. In the
presence of a sequential fourth family of fermions with very high
masses, a Higgs boson with standard model couplings and a mass between 144 and
207 $\GeVcc$ has been excluded at 95\% confidence level.

\section*{Acknowledgments}
We wish to congratulate our colleagues in the CERN accelerator departments for the excellent performance of the
LHC machine. We thank the technical and administrative staff at CERN and other CMS institutes, and acknowledge
support from: FMSR (Austria); FNRS and FWO (Belgium); CNPq, CAPES, FAPERJ, and FAPESP (Brazil); MES (Bulgaria);
CERN; CAS, MoST, and NSFC (China); COLCIENCIAS (Colombia); MSES (Croatia); RPF (Cyprus); Academy of Sciences
and NICPB (Estonia); Academy of Finland, ME, and HIP (Finland); CEA and CNRS/IN2P3 (France); BMBF, DFG, and HGF
(Germany); GSRT (Greece); OTKA and NKTH (Hungary); DAE and DST (India); IPM (Iran); SFI (Ireland); INFN
(Italy); NRF and WCU (Korea); LAS (Lithuania); CINVESTAV, CONACYT, SEP, and UASLP-FAI (Mexico); PAEC
(Pakistan); SCSR (Poland); FCT (Portugal); JINR (Armenia, Belarus, Georgia, Ukraine, Uzbekistan); MST and MAE
(Russia); MSTD (Serbia); MICINN and CPAN (Spain); Swiss Funding Agencies (Switzerland); NSC (Taipei); TUBITAK
and TAEK (Turkey); STFC (United Kingdom); DOE and NSF (USA).
Individuals have received support from the Marie-Curie programme and the European Research Council (European Union);
the Leventis Foundation; the A. P. Sloan Foundation; the Alexander von Humboldt Foundation; the Associazione per
lo Sviluppo Scientifico e Tecnologico del Piemonte (Italy); the Belgian Federal Science Policy Office; the Fonds
pour la Formation \`a la Recherche dans l'Industrie et dans l'Agriculture (FRIA-Belgium); and the Agentschap voor
Innovatie door Wetenschap en Technologie (IWT-Belgium).
\bibliography{auto_generated}   
\cleardoublepage\appendix\section{The CMS Collaboration \label{app:collab}}\begin{sloppypar}\hyphenpenalty=5000\widowpenalty=500\clubpenalty=5000\textbf{Yerevan Physics Institute,  Yerevan,  Armenia}\\*[0pt]
S.~Chatrchyan, V.~Khachatryan, A.M.~Sirunyan, A.~Tumasyan
\vskip\cmsinstskip
\textbf{Institut f\"{u}r Hochenergiephysik der OeAW,  Wien,  Austria}\\*[0pt]
W.~Adam, T.~Bergauer, M.~Dragicevic, J.~Er\"{o}, C.~Fabjan, M.~Friedl, R.~Fr\"{u}hwirth, V.M.~Ghete, J.~Hammer\cmsAuthorMark{1}, S.~H\"{a}nsel, M.~Hoch, N.~H\"{o}rmann, J.~Hrubec, M.~Jeitler, G.~Kasieczka, W.~Kiesenhofer, M.~Krammer, D.~Liko, I.~Mikulec, M.~Pernicka, H.~Rohringer, R.~Sch\"{o}fbeck, J.~Strauss, F.~Teischinger, P.~Wagner, W.~Waltenberger, G.~Walzel, E.~Widl, C.-E.~Wulz
% ...
\vskip\cmsinstskip
% ...
1:~~Also at CERN, European Organization for Nuclear Research, Geneva, Switzerland\\
% ...

\end{sloppypar}
\end{document}